\documentclass[nojss]{jss}
\usepackage[utf8]{inputenc}
\usepackage[T1]{fontenc}
\usepackage{mathtools,bm,subfig}
\pdfoutput=1

\author{Douglas De Rizzo Meneghetti\\Centro Universitário FEI \And
    Plinio Thomaz Aquino Junior\\Centro Universitário FEI}
\title{Application and Simulation of Computerized Adaptive Tests Through the Package \pkg{catsim}}

\Plainauthor{Douglas De Rizzo Meneghetti, Plinio Thomaz Aquino Junior} 
\Plaintitle{Application and Simulation of Computerized Adaptive Tests Through the Package catsim} 
\Shorttitle{\pkg{catsim}: Computerized Adaptive Testing Simulator} 

\Abstract{
    This paper presents \pkg{catsim}, the first package written in the \proglang{Python} language specialized in computerized adaptive tests and the logistical models of Item Response Theory. \pkg{catsim} provides functions for generating item and examinee parameters, simulating tests and plotting results, as well as enabling end users to create new procedures for proficiency initialization, item selection, proficiency estimation and test stopping criteria. The simulator keeps a record of the items selected for each examinee as well as their answers and also enables the simulation of linear tests, in which all examinees answer the same items. The various components made available by \pkg{catsim} can also be used in the creation of third-party testing applications. Examples of such usages are also presented in this paper.
}
\Keywords{computerized adaptive testing, item response theory, \proglang{Python}}
\Plainkeywords{computerized adaptive testing, item response theory, Python} 

\Address{
    Department of Graduate Electrical Engineering\\
    Centro Universitário FEI\\
    3972 Humberto de Alencar Castelo Branco Avenue, Sao Paulo, Brazil\\

    Douglas De Rizzo Meneghetti\\
    E-mail: \email{douglasrizzo@fei.edu.br}\\
    URL: \url{http://douglasrizzo.com.br/}\\

    Plinio Thomaz Aquino Junior\\
    E-mail: \email{plinio.aquino@fei.edu.br}\\
    URL: \url{http://fei.edu.br/~plinio.aquino/}
}

\DeclareMathOperator*{\argmax}{arg\,max}

\begin{document}



\section{Introduction}

Assessment instruments are widely used to measure individuals \emph{latent traits}, that is, internal characteristics that cannot be directly measured. An example of such assessment instruments are educational and psychological tests. Each test is composed of a series of items and an examinee's answers to these items allow for the measurement of one or more of their latent traits. When a latent trait is expressed in numerical form, it is called an \emph{ability} or \emph{proficiency}.

Ordinary tests, hereon called linear tests, are applied using the orthodox paper and pencil strategy, in which tests are printed and all examinees are presented with the same items in the same order. One of the drawbacks of this methodology is that individuals with either high or low proficiencies must answer all items in order to have their proficiency estimated. An individual with high proficiency might get bored of answering the whole test if it only contains items that he or she considers easy; on the other hand, an individual of low proficiency might get frustrated if he is confronted by items considered hard and might give up on the test or answers the items without paying attention.

With these concerns in mind, a new paradigm in assessment emerged in the 70s. Initially named \emph{tailored testing} by \citet{lord_statistical_1968}, these were tests in which items were chosen to be presented to the examinee in real time, based on the examinee's responses to previous items. The name was changed to computerized adaptive testing (CAT) due to the advances in technology that facilitated the application of such a testing methodology using electronic devices, like computers and tablets.

In a CAT, the examinee's proficiency is evaluated after the response of each item. The new proficiency is then used to select a new item, theoretically closer to the examinee's real proficiency. This method of test application has several advantages compared to the traditional paper-and-pencil method, since examinees of high proficiency are not required to answer all the low difficulty items in a test, answering only the items that actually give high information regarding his or hers true knowledge of the subject at matter. A similar, but inverse effect happens for those examinees of low proficiency level.

Finally, the advent of CAT allowed for researchers to create their own variant ways of starting a test, choosing items, estimating proficiencies and stopping the test. Fortunately, the mathematical formalization provided by Item Response Theory (IRT) allows for tests to be computationally simulated and the different methodologies of applying a CAT to be compared under different constraints. Packages with these functionalities already exist in the \proglang{R} language \citep[cf. \citeauthor{magis_random_2012}, \citeyear{magis_random_2012}]{r_language_2018} but not yet in \proglang{Python}, to the best of the authors' knowledge. \pkg{catsim} was created to fill this gap, using the facilities of established scientific packages such as \pkg{numpy} \citep{oliphant_guide_2015} and \pkg{scipy} \citep{jones_scipy_2001}, as well as the object-oriented programming paradigm supported by \proglang{Python} to create a simple, comprehensive and user-extendable CAT simulation package.

Lastly, it is important to inform the reader that \pkg{catsim} has been aimed towards \proglang{Python} 3 from its advent\footnote{See \url{https://python3statement.org/} for a list of projects that propose to step down \proglang{Python} 2.7 support until 2020.}, in order to simplify its development process and make use of features available only in the new version of the language.

\section{Related Work}

The work on \pkg{catsim} is comparable to similar open-source CAT-related packages such as \pkg{catR} \citep{magis_random_2012}, \pkg{catIrt} \citep{steven_w._nydick_catirt:_2015} and \pkg{MAT} \citep{seung_w._choi_mat:_2015}. While \pkg{catsim} does not yet implement Bayesian examinee proficiency estimation (like the three aforementioned packages) and simulation of tests under Multidimensional Item Response Theory (like \pkg{MAT}), it contributes with a range of item selection methods from the literature, as well as by taking a broader approach by providing users with IRT-related functions, such as bias, item and test information, as well as functions that assist plotting item curves and test results. Furthermore, \pkg{catsim} is the only package of its kind developed in the \proglang{Python} language, to the best of the authors' knowledge.

Similarly, another open-source package that works with adaptive testing is \pkg{mirtCAT} \citep{chalmers_generating_2016}, which provides users an interface that enables the creation of both adaptive and non-adaptive tests. While \pkg{catsim}'s main purpose is not the presentation of tests, it allows users to both simulate tests under different circumstances and anticipate possible issues before creating a full-scale test as well as use its components in the development of software capable of administering a CAT.

CATSim, a homonymous program by Assessment Systems\footnote{\url{http://www.assess.com/catsim}}, is a commercial CAT analysis tool created with the purpose of assisting in the development of adaptive testing. While it provides further constraints on item selection as well as the inclusion of polytomous models, it is supposed to be used as a standalone application and not as an extensible software library and thus does not allow users to create their own item selection methods (constrained or otherwise) as well as integrate with other software as \pkg{catsim} allows.

\section{Item Response Theory}

As a CAT simulator, \pkg{catsim} borrows many concepts from Item Response Theory (IRT) \citep{lord_statistical_1968, rasch_item_1966}, a series of models created in the second part of the 20th century with the goal of \emph{measuring latent traits}. \pkg{catsim} makes use of Item Response Theory one-, two-, three- and four-parameter logistic models, in which examinees and items are represented by a set of numerical values (the models' parameters). Item Response Theory itself was created with the goal of measuring latent traits as well as assessing and comparing individuals' proficiencies by allocating them in proficiency scales, inspiring as well as justifying its use in adaptive testing.

In the unidimensional logistic models of Item Response Theory, a given assessment instrument only measures a single proficiency (or dimension of knowledge). The instrument, in turn, is composed of \emph{items} in which examinees manifest their latent traits when answering them. 

In unidimensional IRT models, an examinee's proficiency is represented as \(\theta\). Usually, \(-\inf < \theta < \inf\), but since the scale of \(\theta\) is defined by the individuals creating the instrument, it is common for the values to be fit to a normal distribution \(N(0; 1)\), such that \(-4 < \theta < 4\). Additionally, \(\hat\theta\) is the estimate of \(\theta\). Since a latent trait can't be measured directly, estimates need to be made, which tend to get closer to the theoretically real \(\theta\) as the test progresses in length.

Under the models of IRT, each item is represented by a set of parameters. The four-parameter logistic model represents each item using the following parameters:

\begin{itemize}
	\item \(a\) represents an item's \emph{discrimination} parameter, that is, how well it discriminates individuals who answer the item correctly (or, in an alternative interpretation, individuals who agree with the idea of the item) and those who don't. An item with a high \(a\) value tends to be answered correctly by all individuals whose \(\theta\) is above the items difficulty level and wrongly by all the others; as this value gets lower, this threshold gets blurry and the item starts not to be as informative. It is common for \(a > 0\);
	\item \(b\) represents an item's \emph{difficulty} parameter. This parameter, which is measured in the same scale as \(\theta\), denotes the point where \(P(X=1)=0.5\). For a CAT, it is good for an item bank to have as many items as possible in all difficulty levels, so that the CAT may select the best item for each individual in all ability levels;
	\item  \(c\) represents an item's \emph{pseudo-guessing} parameter. This parameter denotes the probability of individuals with extremely low proficiency values to still answer the item correctly. Since \(c\) is a probability, \(0 < c \leq 1\). The lower the value of this parameter, the better the item is considered;
	\item  \(d\) represents an item's \emph{upper asymptote}. This parameter denotes the probability of individuals with extremely high proficiency values to still answer the item incorrectly. Since \(d\) represents an upper limit of a probability function, \(0 < d \leq 1\). The lower the value of this parameter, the better the item is considered.
\end{itemize}

In an item bank with \(N\) items, when \(\forall i \in N, d_i = 1\), the four-parameter logistic model is reduced to the three-parameter logistic model; when \(\forall i \in N, c_i = 0\), the three-parameter logistic model is reduced to the two-parameter logistic model; if all values of \(a\) are equal, the two-parameter logistic model is reduced to the one-parameter logistic model. Finally, when \(\forall i \in N, a_i = 1\), we have the Rasch model \citep{rasch_item_1966}. Thus, \pkg{catsim} is able of treating all of the logistic models presented above, since the underlying functions of all logistic models related to test simulations are the same, given the correct item parameters.

Under the four-parameter logistic model of IRT, the probability of an examinee with a given \(\theta\) value to answer item \(i\) correctly, given the item parameters, is given by \citep{magis_note_2013}

\begin{equation}\label{eq:icc}
	P(X_i = 1 | \theta) = c_i + \frac{d_i-c_i}{1+ e^{a_i(\theta-b_i)}}
\end{equation}

The information of item \(i\) is calculated as \citep{magis_note_2013}

\begin{equation}\label{eq:iif}
	I_i(\theta) = \frac{a_i^2[(P(\theta)-c_i)]^2[d_i - P(\theta)]^2}{(d_i - c_i)^2(1-P(\theta))P(\theta)},
\end{equation}

where \(P(\theta)\) is used as a shorthand for \(P(X=1|\theta)\).

In the one-parameter and two-parameter logistic models, an item \(i\) is most informative when its difficulty parameter is close to the examinee's proficiency, that is, \(b_i = \theta\). In the three-parameter and four-parameter logistic models, the \(\theta\) value where each item maximizes its information is given by \citep{magis_note_2013}

\begin{equation}\label{eq:maxinf}
	\argmax_{\theta}I(\theta) = b + \frac{1}{a} \log{\left(\frac{x^* - c}{d - x^*}\right)}
\end{equation}

where

\[ x^* = 2 \sqrt{\frac{-u}{3}} cos\left\{\frac{1}{3}acos\left(-\frac{v}{2}\sqrt{\frac{27}{-u^3}}\right)+\frac{4 \pi}{3}\right\} + 0.5 \]

\[ u = -\frac{3}{4} + \frac{c + d - 2cd}{2} \]

\[ v = -\frac{c + d - 1}{4} \]

\begin{figure}
	\centering
	\includegraphics[width=.6\textwidth]{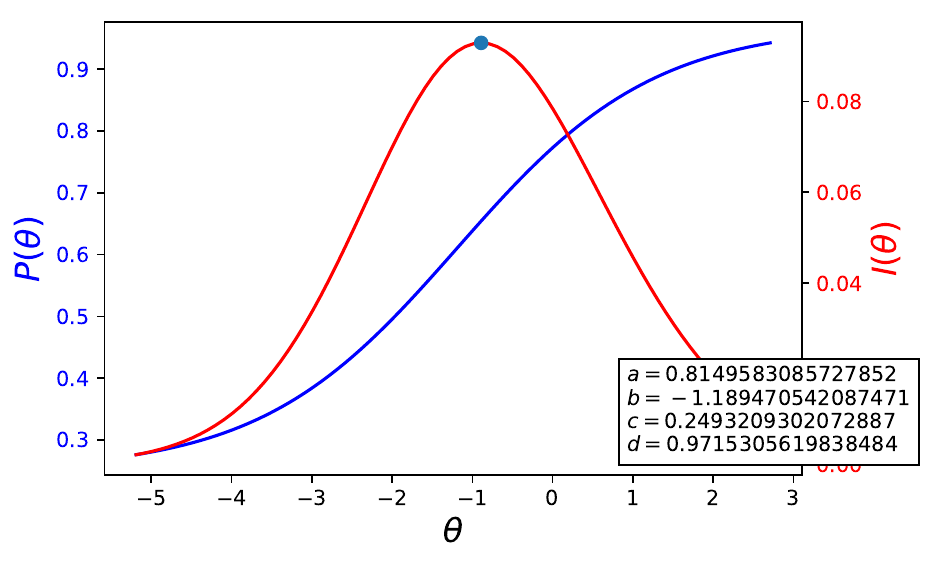}
	\caption{An item characteristic curve and its corresponding information curve, created with \pkg{catsim} plotting functions.}\label{fig:item_curve}
\end{figure}

The item characteristic curve, given by Equation~\ref{eq:icc}; the item information curve, given by Equation~\ref{eq:iif}; and the maximum information point, obtained with Equation~\ref{eq:maxinf} are graphically represented in figure \ref{fig:item_curve} for a randomly generated item.

Equations \ref{eq:icc}, \ref{eq:iif} and \ref{eq:maxinf} are available in \pkg{catsim} under module \code{catsim.irt} as \code{icc()}, \code{inf()} and \code{max_info()}, respectively. High performance versions of the same functions, which return the response probability and information for all items in the bank, given a \(\theta\) value, or return the \(\theta\) values that maximize information for the entire item bank at once, are also available in the same module as \code{icc_hpc()}, \code{inf_hpc()} and \code{max_info_hpc()}. These versions were developed using the \pkg{numexpr} \citep{numexpr} package.

The sum of the information of all \(J\) items in a test is called \emph{test information} \citep{de_ayala_theory_2009}:

\[I(\theta) = \sum_{i \in J} I_i(\theta).\]

The amount of error in the estimate of an examinee's proficiency after a test is called the \emph{standard error of estimation} \citep{de_ayala_theory_2009} and it is given by

\begin{equation} \label{eq:see}
	SEE = \sqrt{\frac{1}{I(\hat\theta)}}.
\end{equation}

Since the denominator in the right-hand side of the \(SEE\) is \(I(\hat\theta)\), it is clear to see that the more items an examinee answers, the smaller \(SEE\) gets. Additionally, the variance in the estimation of \(\theta\) is given by \(Var = SEE^2\).

Another measure of internal consistency for the test is test reliability, a value with similar applicability to Cronbach's \(\alpha\) in Classical Test Theory. It is given by \citep{thissen2000reliability} \[ Rel = 1 - \frac{1}{I(\hat\theta)}. \] Its value is always lower than 1, with values close to 1 indicating good reliability. Note, however, that when \(I(\theta) < 1\), \(Rel < 0\). In such cases, the use of \(Rel\) does not make sense, but this situation may be fixed with the application of additional items to the examinee.

\pkg{catsim} provides these functions in the \code{catsim.irt} module.

\subsection{Item bank}

In \pkg{catsim}, a collection of items is represented as a \code{numpy.ndarray} whose rows and columns represent items and their parameters, respectively. Thus, it is referred to as the \emph{item matrix}. Item parameters \(a\), \(b\), \(c\) and \(d\) are situated in the first four columns of the matrix, respectively.

Users can create their own item matrices as \code{numpy.ndarray}s. If the array contains a single column, it will be treated a the difficulty parameter. In an array with two columns, the second column is assumed to be the discrimination parameters of each item. With three columns, the third column is treated as the pseudo-guessing parameter. Finally, when all four columns are passed, the last column is assumed to be the upper asymptote of each item.

\pkg{catsim} provides functions to validate and normalize item matrices by checking if item parameters are within their minimum and maximum bounds and creating additional columns in the matrix. \(1 \times N\), \(2 \times N\) and \(3 \times N\) matrices can be transformed into a \(4 \times N\) matrix using the \code{irt.normalize_item_bank()} function, which assumes existent columns represent parameters \(b\), \(a\), \(c\) and \(d\), in that order, and adds any missing columns with default values, in such a way that the matrix still represents items in their original logistic models.

If any simulations are ran using a given item matrix, \pkg{catsim} adds a fifth column to the matrix, containing each item's exposure rate, commonly denoted as \(r\). Its value denotes the ratio of tests in which item \(i\) has appeared and it is calculated as follows:

\[r_i = \frac{M_i}{M}\]

Where \(M\) represents the total number of tests applied and \(M_i\) is the number of tests item \(i\) has appeared on.

Finally, item matrices can be generated via the \code{catsim.cat.generate_item_bank} function as follows:

\begin{verbatim}
items = generate_item_bank(5, '1PL')
items = generate_item_bank(5, '2PL')
items = generate_item_bank(5, '3PL')
items = generate_item_bank(5, '4PL')
items = generate_item_bank(5, '4PL', corr=0.5)
\end{verbatim}

These examples depict the generation of an array of five items according to the different logistic models. In the last example, parameters \(a\) and \(b\) have a correlation of \(0.5\), an adjustment that may be useful in case simulations require it \citep{chang_-stratified_2001}. Parameters are extracted from the following probability distributions: \(a \sim N(1.2, 0.25)\), \(b \sim N(0, 1)\), \(c \sim \max(N(0.25, 0.02), 0)\) and \(d \sim U(0.94, 1)\), generated using \pkg{numpy}'s \code{random} module. These distributions are assumed to follow simple real world distributions \citep{barrada_method_2010}.

\section{Computerized adaptive testing}

Unlike linear tests, in which items are sequentially presented to examinees and their proficiency estimated at the end of the test, in a computerized adaptive test (CAT), an examinees' proficiency is calculated after the response of each item \citep{lord_broad-range_1977,lord_applications_1980}. The updated knowledge of an examinee's proficiency at each step of the test allows for the selection of more informative items \emph{during} the test itself, which in turn reduce the standard error of estimation of their proficiency at a faster rate.

\begin{figure}
	\centering
	\includegraphics[width=.5\textwidth]{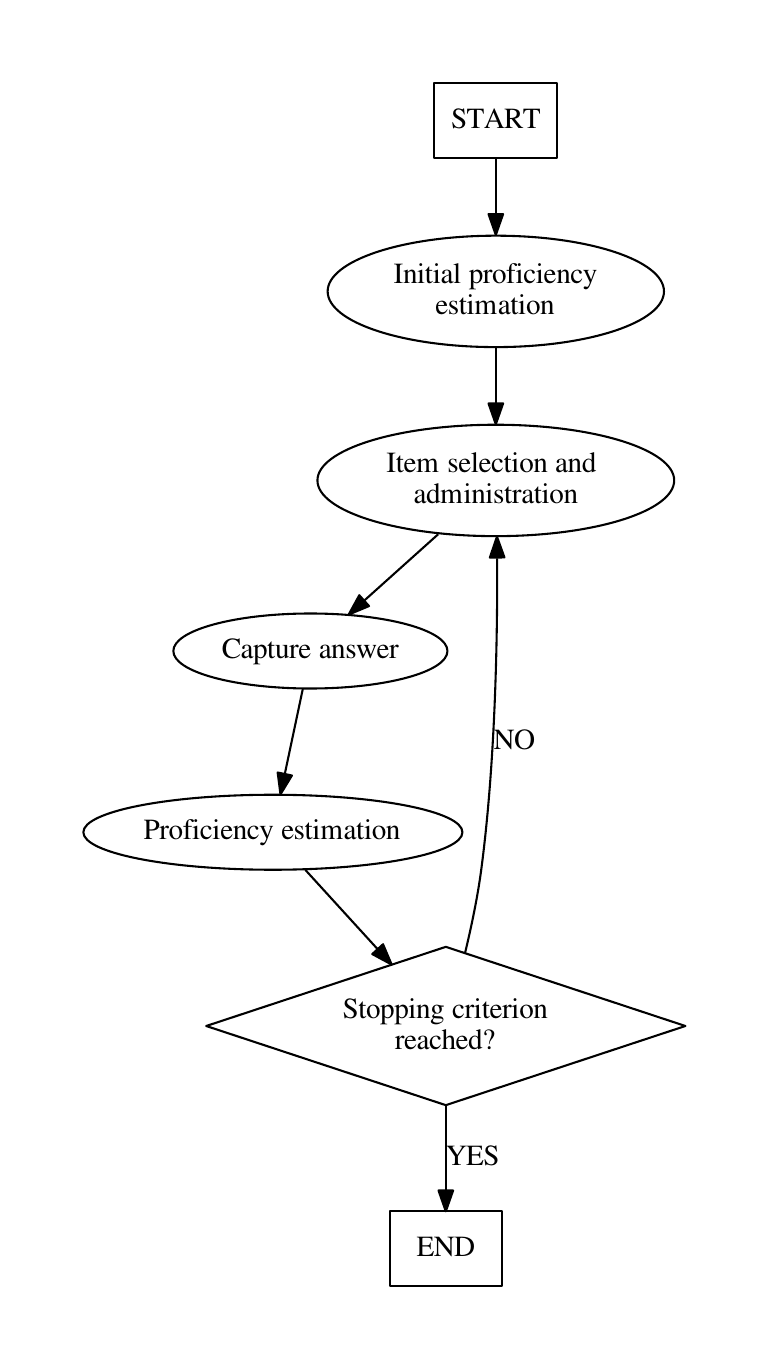}
	\caption{The iterative process of a computerized adaptive test.}\label{fig:cat}
\end{figure}

In general, a computerized adaptive test has a very well-defined lifecycle, presented in figure \ref{fig:cat}.

\begin{enumerate}
	\item The examinee's initial proficiency is estimated;
	\item \label{itm:two} An item is selected based on the current proficiency estimation;
	\item The proficiency is reestimated based on the answers to all items up until now;
	\item \textbf{If} a stopping criterion is met, stop the test. \textbf{Else} go back to step \ref{itm:two}.
\end{enumerate}

In \pkg{catsim}, each phase is separated in its own module, which makes it easy to create simulations combining different methods for each phase. Each module will be explained separately.

\section{Package architecture}

\pkg{catsim} was built using an object-oriented architecture, an approach which introduces benefits for the package maintenance and expansion. Each phase in the CAT lifecycle is represented by a different abstract class in the package. Each of these classes represent either a proficiency initialization procedure, an item selection method, a proficiency estimation procedure or a stopping criterion. The implementations of these procedures are made available in dedicated modules. Users and contributors may independently implement their own methods for each phase of the CAT lifecycle, or even an entire new CAT iterative process while still using \pkg{catsim} and its features to simulate tests and plot results, they respect the contracts of the provided abstract classes. Modules and their corresponding abstract classes are presented on table \ref{tbl:modules_classes}.

\begin{table}
	\centering
	\begin{tabular}{|r|l|}
		\hline
		\textbf{Abstract class}              & \textbf{Module with implementations} \\
		\hline
		\code{catsim.simulation.Initializer} & \code{catsim.initialization}         \\
		\hline
		\code{catsim.simulation.Selector}    & \code{catsim.selection}              \\
		\hline
		\code{catsim.simulation.Estimator}   & \code{catsim.estimation}             \\
		\hline
		\code{catsim.simulation.Stopper}     & \code{catsim.stopping}               \\
		\hline
	\end{tabular}
	\caption{Modules involved in the simulation and their corresponding abstract classes.}\label{tbl:modules_classes}
\end{table}

\subsection{Proficiency initialization}

The initialization procedure is done only once during each examinee's test. In it, the initial value of an examinee's proficiency \(\hat\theta_0\) is selected. Ideally, the closer \(\hat\theta_0\) is to \(\theta\), the faster it converges to \(\theta\), the examinee's true proficiency value \citep{sukamolson_computerized_2002}. The initialization procedure may be done in a variety of ways. A standard value can be chosen to initialize all examinees (\code{FixedInitializer}); it can be chosen randomly from a probability distribution (\code{RandomInitializer}); the place in the item bank with items of more information can be chosen to initialize \(\hat\theta_0\) \citep{dodd_effect_1990} etc.

In \pkg{catsim}, initialization procedures can be found in the \code{catsim.initialization} module.

\subsection{Item selection}

During a test in which an examinee has already answered \(t\) items, the answers to all \(t\) items are used to estimate \(\hat\theta_{t}\), which may be used to select item \(t+1\) or to stop the test, if a stopping criteion has been met.

Table \ref{tbl:sel_methods} lists the available selection methods provided by \pkg{catsim}. Selection methods can be found in the \code{catsim.selection} module.

\begin{table}
	\centering
	\begin{tabular}{r|l}
		\textbf{Class}                       & \textbf{Author}                     \\
		\hline
		\code{AStratifiedBBlockingSelector}  & \citet{chang_-stratified_2001}      \\
		\code{AStratifiedSelector}           & \citet{chang_-stratified_1999}      \\
		\code{ClusterSelector}               & \citet{meneghetti_metolodogia_2015} \\
		\code{IntervalIntegrationSelector}   & ---                                 \\
		\code{LinearSelector}                & ---                                 \\
		\code{MaxInfoBBlockingSelector}      & \citet{barrada_maximum_2006}        \\
		\code{MaxInfoSelector}               & \citet{lord_broad-range_1977}       \\
		\code{UrrySelector}                  & \citet{urry_monte_1970}             \\
		\code{MaxInfoStratificationSelector} & \citet{barrada_maximum_2006}        \\
		\code{RandomSelector}                & ---                                 \\
		\code{RandomesqueSelector}           & \citet{kingsbury_procedures_1989}   \\
		\code{The54321Selector}              & \citet{mcbride_reliability_1983}    \\
		\hline
	\end{tabular}
	\caption{Item selection methods implemented in \pkg{catsim}.}\label{tbl:sel_methods}
\end{table}

\code{MaxInfoSelector} selects \(i_{t+1}\) as the item that maximizes information gain, that is, to select \(\argmax_{i \in N} I_i(\hat\theta_t)\). This method guarantees faster decrease of standard error. However, it has been shown to have drawbacks, like causing overexposure of few items in the item bank while ignoring items with smaller information values.

\code{AStratifiedSelector}, \code{MaxInfoStratificationSelector}, \code{AStratifiedBBlockingSelector} and \code{MaxInfoBBlockingSelector} are classified \citep{georgiadou_review_2007} as stratification methods. In the \code{AStratifiedSelector} selector, the item bank is sorted in ascending order according to the items discrimination parameter and then separated into \(K\) strata (\(K\) being the test size), each stratum containing gradually higher average discrimination. The \(\\alpha\)-stratified selector then selects the first non-administered item from stratum \(k\), in which \(k\) represents the position in the test of the current item the examinee is being presented. \code{MaxInfoStratificationSelector} separates items in strata according to their maximum \(I(\theta)\) value. \code{AStratifiedBBlockingSelector} and \code{MaxInfoBBlockingSelector} are variants in which item difficulty is guaranteed to have a distribution close to uniform among the \(K\) strata, after the discovery that there is a positive correlation between \(a\) and \(b\) \citep{chang_-stratified_2001}.

\code{RandomSelector}, \code{RandomesqueSelector} and \code{The54321Selector} are methods that apply randomness to item selection. \code{RandomSelector} selects items randomly from the item bank; \code{RandomesqueSelector} selects an item randomly from the \(n\) most informative items at that step of the test, \(n\) being a user-defined variable. \code{The54321Selector} also selects between the \(n\) most informative items, but at each new item that is selected, \(n=n-1\), guaranteeing that more informative items are selected as the test progresses.

\code{LinearSelector} is a selector that mimics a linear test, that is, item indices are predefined and all examinees are presented the same items. This method is useful for the simulation of an ordinary paper and pencil test.

\code{ClusterSelector} clusters items according to their parameter values and selects items from the cluster that contains either the most informative item or the highest average information.

\code{IntervalIntegrationSelector} is an experimental selection method that selects the item whose integral around an interval \(\delta\) of the peak of information is maximum, like so \[ \argmax_{i \in N} \int_{\hat\theta - \delta}^{\hat\theta + \delta}I_i(\hat\theta). \] It uses \pkg{scipy.integrate} \code{quad} function for the process of integration.

\subsection{Proficiency estimation}

Proficiency estimation occurs whenever an examinee answers a new item. After an item is selected, \pkg{catsim} simulates an examinee's response to a given item by sampling a binary value from the Bernoulli distribution, in which the value of \(p\) is given by Equation~\ref{eq:icc}. Given a dichotomous (binary) response vector and the parameters of the corresponding items that were answered, it is the job of an estimator to return a new value for the examinee's \(\hat\theta\). This value reflects the examinee's proficiency, given his or hers answers up until that point of the test.

In \proglang{Python}, an example of a list that may be used as a valid dichotomous response vector is as follows:

\code{response_vector = [True, True, True, False, True, False, True, False, False]}

Estimation techniques are generally separated between maximum-likelihood estimation procedures (whose job is to return the \(\hat\theta\) value that maximizes a \emph{likelihood} function, presented in \code{catsim.irt.log_likelihood}) and Bayesian estimation procedures, which use prior information of the distributions of examinee's proficiencies to estimate new values for them.

The likelihood of an examinee answering a set of \(J\) items with response pattern \(\bm{X_J}\), given a value of \(\hat\theta\) and the parameters \(\bm{\zeta_J}\) of the \(J\) items is given by the following likelihood function

\[L(\bm{X_J} | \theta, \bm{\zeta_J}) = \prod_{i=1} ^ J P_{i}(\theta)^{X_{i}} Q_{i}(\theta)^{1-X_{i}}\]

For mathematical reasons, finding the maximum of \(L)\) includes using the product rule of derivations. Since \(L)\) has \(i\) parts, it can be quite complicated to do so. Also, for computational reasons, the product of probabilities can quickly tend to 0, so it is common to use the log-likelihood in maximization/minimization problems, transforming the product of probabilities in a sum of probabilities. This function, presented in \pkg{catsim} as \code{catsim.irt.log_likelihood}, is given by

\[\log L(\bm{X_J} | \theta, \bm{\zeta_J}) = \sum_{i=1} ^ J \left\lbrace X_{i} \log P_{i}(\theta)+ (1 - X_{i}) \log Q_{i}(\theta) \right\rbrace .\]

\pkg{catsim} provides a maximum likelihood estimation procedure based on the hill climbing optimization technique, called \code{HillClimbingEstimator}. Since maximum likelihood estimators tend to fail when all observed values are equal, the provided estimation procedure employs the method proposed by \citet{dodd_effect_1990} to select the value of \(\hat\theta\) as follows:

\[ \hat{\theta}_{t+1} = \left\lbrace \begin{array}{ll}
		\hat{\theta}_t+\frac{b_{max}-\hat{\theta_t}}{2} & \text{if } X_t = 1  \\
		\hat{\theta}_t-\frac{\hat{\theta}_t-b_{min}}{2} & \text{if }  X_t = 0
	\end{array} \right\rbrace , \]

where \(b_{max}\) and \(b_{min}\) are the maximum and minimum \(b\) values in the item bank, respectively.

The second estimation method uses \pkg{scipy} \citep{jones_scipy_2001} optimization module, more specifically the \code{differential_evolution()} function, to find the global maximum of the likelihood function. These proficiency estimation procedures can be found in the \code{catsim.estimation} module.

\begin{figure}[!ht]
	\centering
	\subfloat[\code{DifferentialEvolutionEstimator} (38 avg. evals)]{\includegraphics[width= 0.49\columnwidth]{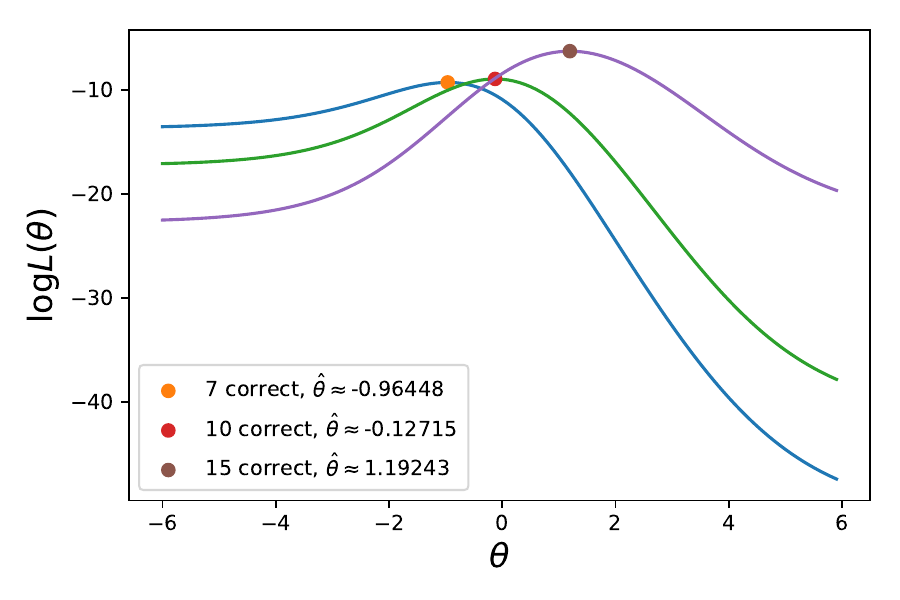}}
	\subfloat[\code{HillClimbingEstimator} (8 avg. evals)]{\includegraphics[width= 0.49\columnwidth]{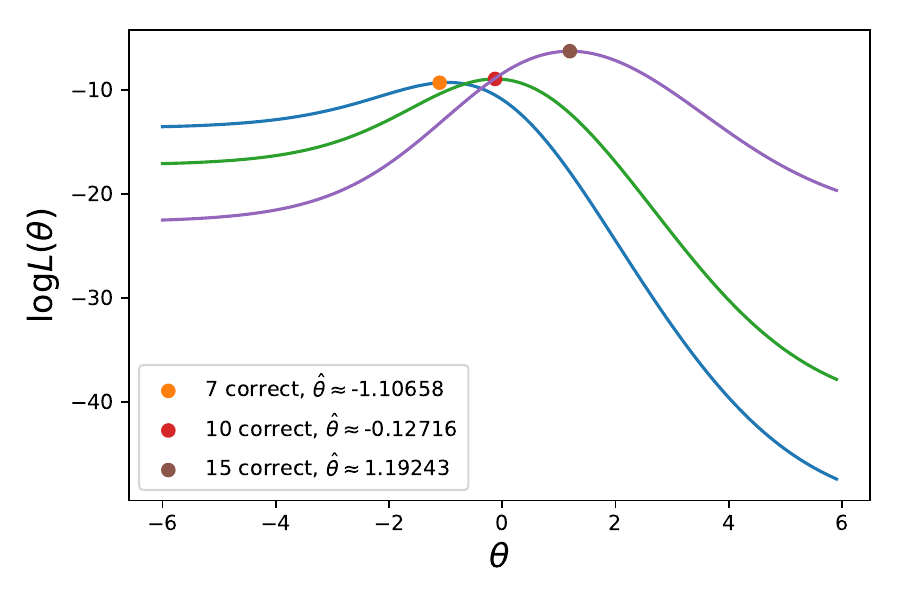}}
	\caption{Average number of function evaluations and graphical representation of function maxima found by the available estimation methods in \pkg{catsim}.}\label{fig:est_eval}
\end{figure}

Figure \ref{fig:est_eval} shows the performance of the available estimators. \code{HillClimbingEstimator} evaluates the log-likelihood function an average of 7 times before reaching a step-size smaller than \(10e-5\), while the \pkg{scipy}-based \code{DifferentialEvolutionEstimator} evaluates the function four times more.

Estimators may benefit from prior knowledge of the estimate, such as known lower and upper bounds for \(\hat\theta\) or the value of \(\hat\theta_t\) when estimating \(\hat\theta_{t + 1}\).

\subsection{Stopping criterion}

Since items in a CAT are selected during the test, a stopping criterion must be chosen such that, when achieved, no new items are presented to the examinee and the test is deemed finished. These stopping criteria might be achieved when the test reaches a fixed number of items or when the standard error of estimation from Equation~\ref{eq:see} \citep{de_ayala_theory_2009} reaches a lower threshold \citep{lord_broad-range_1977}, among others. Both of the aforementioned stopping criteria are implemented as \code{MaxItemStopper} and \code{MinErrorStopper}, respectively, and are located in the \code{catsim.stopping} module.

\section{Usage in simulations}

The simulation process is configured using the \code{Simulator} class from the \code{catsim.simulation} module. A \code{Simulator} is created by passing an \code{Initializer}, a \code{Selector}, an \code{Estimator} and a \code{Stopper} object to it, as well as either an empirical or generated item matrix and a list containing examinees \(\theta\) values or a number of examinees. If a number of examinees is passed, \pkg{catsim} generates their \(\theta\) values from the normal distribution, using \(\mu_\theta = \mu_b\) and \(\sigma_\theta = \sigma_b\), where \(b\) represents the difficulty parameter of all items in the item matrix. If an item matrix is not provided, \(\mu_\theta = 0\) and \(\sigma_\theta = 1\).

When used in conjunction with the \code{Simulator}, all four abstract classes (\code{Initializer}, \code{Selector}, \code{Estimator} and \code{Stopper}) as well as their implementations have complete access to the current and past states of the simulation, which include:

\begin{enumerate}
	\item item parameters;
	\item indexes of the administered items to all examinees;
	\item current and past proficiency estimations of all examinees;
	\item each examinee's response vector.
\end{enumerate}

The following example depicts the process of creating the necessary objects to configure a \code{Simulator} and calling the \code{simulate()} method to start the simulation.

\begin{verbatim}
from catsim.cat import generate_item_bank
from catsim.initialization import RandomInitializer
from catsim.selection import MaxInfoSelector
from catsim.estimation import HillClimbingEstimator
from catsim.stopping import MaxItemStopper
from catsim.simulation import Simulator

initializer = RandomInitializer()
selector = MaxInfoSelector()
estimator = HillClimbingEstimator()
stopper = MaxItemStopper(20)
s = Simulator(generate_item_bank(100), 10)
s.simulate(initializer, selector, estimator, stopper)
\end{verbatim}

During the simulation, the \code{Simulator} calculates and maintains \(\hat\theta\), \(SEE\) and \(Var\) after every item is presented to each examinee, as well as the indexes of the presented items in the order they were presented. These values are exposed as properties of the \code{Simulator} and are accessed as \code{s.estimations}, \code{s.see}, \code{s.var}, \code{s.administered_items} and so on.

\section{Integration with other applications}

Aside from being used in the simulation process, implementations of \code{Initializer}, \code{Selector}, \code{Estimator} and \code{Stopper} may also be used independently. Given the appropriate parameters, each component is able to return results in an accessible way, allowing \pkg{catsim} to be used in other software programs. Table \ref{tbl:independent} shows the main functions of each component, the parameters they commonly use and their return values.

\begin{table}
	\centering
	\begin{tabular}{|r|c|p{5.5cm}|p{3cm}|}
		\hline
		\textbf{Class}     & \textbf{Main method} & \textbf{Common parameters}                                                  & \textbf{Returns}                                     \\
		\hline
		\code{Initializer} & \code{initialize}    & None                                                                        & \(\hat\theta_0\)                                     \\
		\hline
		\code{Selector}    & \code{select}        & item bank, indexes of administered items, \(\hat\theta_t\)                  & Index of the next item                               \\
		\hline
		\code{Estimator}   & \code{estimate}      & item bank, indexes of administered items, \(\hat\theta_t\), response vector & \(\hat\theta_{t+1}\)                                 \\
		\hline
		\code{Stopper}     & \code{stop}          & parameters of administered items, \(\hat\theta_t\)                          & \code{True} to stop the test, otherwise \code{False} \\
		\hline
	\end{tabular}
	\caption{Main functions of each component and the parameters they commonly use.}\label{tbl:independent}
\end{table}

The code below exemplifies the independent usage of each CAT component outside of the simulator. We'll first generate an item matrix, a response vector and a list containing the indices of administered items to a given examinee.

\begin{verbatim}
bank_size = 5000
items = generate_item_bank(bank_size)
responses = [True, True, False, False]
administered_items = [1435, 3221, 17, 881]
\end{verbatim}

An \code{Initializer} gives the value of \(\hat\theta_0\).

\begin{verbatim}
initializer = RandomInitializer()
est_theta = initializer.initialize()
\end{verbatim}

Given an item matrix, a list containing the indices of items already administered to an examinee and the response vector to the administered, an \code{Estimator} calculates a new value for \(\hat\theta\). The \code{HillClimbingEstimator} also asks for an initial proficiency estimation, which is necessary for Dodd's item selection procedure \citep{dodd_effect_1990}.

\begin{verbatim}
estimator = HillClimbingEstimator()
new_theta = estimator.estimate(items=items,
    administered_items=administered_items,
    response_vector=responses, est_theta=est_theta)
\end{verbatim}

A \code{Selector} returns the index of the next item to be administered to the current examinee. In the following example, the \code{MaxInfoSelector} needs the item matrix and a \(\hat\theta\) as parameters, so it can return the index of the item that maximizes the information value for that \(\hat\theta\) value. It also uses the indices of administered items to ignore these items during the process.

\begin{verbatim}
selector = MaxInfoSelector()
item_index = selector.select(items=items,
    administered_items=administered_items,
    est_theta=new_theta)
\end{verbatim}

Lastly, a \code{Stopper} returns a boolean value, informing whether the test should be stopped. In the case of the \code{MaxItemStopper}, it returns \code{True} if the number of administered items is greater than or equal to a predetermined value.

\begin{verbatim}
stopper = MaxItemStopper(20)
stop = stopper.stop(administered_items=items[administered_items],
    theta=est_theta)
\end{verbatim}

\section{Validity measures}

A series of validity measures have been implemented, in order to evaluate the results of the adaptive testing simulation. The first two are measurement bias and mean squared error \citep{chang_-stratified_2001}, while the last two are the root mean squared error and test overlap rate \citep{barrada_maximum_2006, barrada_methods_2007, barrada_test_2009, barrada_method_2010, barrada_optimal_2014}.

\[Bias = \frac{\sum_{g=1}^{G} (\hat{\theta}_g - \theta_{g})}{G}\]

\[MSE = \frac{\sum_{g=1}^{G} (\hat{\theta}_g - \theta_{g})^2}{G}\]

\[RMSE = \sqrt{MSE}\]

\[T = \frac{N}{J}S_{\bm{r_N}}^2 + \frac{J}{N}\]

The first three are indicators of the error of measurement between the \(G\) examinees' true proficiency \(\theta_j\) and the proficiency estimated at the end of the CAT, \(\hat\theta_g\). Test overlap rate, according to \citet{barrada_test_2009}, ``provides information about the mean proportion of items shared by two examinees'', where \(N\) indicates the number of items in the item bank, \(J\) is the number of items in the test and \(S_{\bm{r_N}}^2\), the variance in the exposure rates of all items in the item bank. For example, a value of \( T = 0.3\) indicates that two examinees share, on average, \(30\%\) of the same items in a test. It is possible to see that, since \(\frac{J}{N}\) is a constant, \(T\) reaches its lowest value when \(S_{\bm{r_N}}^2 \approx 0\). For that, all items in the item bank must be used in a homogeneous way.

The bias, mean-squared error and root mean-squared error can be calculated as follows:

\begin{verbatim}
cat.bias(true_proficiencies, estimated_proficiencies)
cat.mse(true_proficiencies, estimated_proficiencies)
cat.rmse(true_proficiencies, estimated_proficiencies)
\end{verbatim}

where \code{true_proficiences} and \code{estimated_proficiences} are lists or numpy.ndarrays containing examinees true estimated proficiencies, respectively. The overlap rate can be calculated as follows:

\begin{verbatim}
cat.overlap_rate(items, test_size)
\end{verbatim}

where items is a \code{list} or \code{numpy.ndarray} containing the number of times each item has been used during a series of tests and \code{test_size} is an integer indicating the number of items in the test, assuming the CAT is configured to end with a fixed number of items.

When used in a simulation context, the \code{Simulator} object calculates the aforementioned validity measures and exposes them as properties, which are accessed as \code{bias}, \code{mse}, \code{rmse} and \code{overlap_rate}.

\section{Plotting items and results}

\pkg{catsim} provides a series of plotting functions in the \code{catsim.plot} module, which uses \pkg{matplotlib} \citep{hunter_matplotlib_2007} to plot item curves, as well as plotting proficiency estimations and validity measures for each item answered by every examinee during the simulation.

The following examples shows how to plot an item's characteristic curve, its information curve as well as both curves in the same figure, as presented in figure \ref{fig:item_curve}.

\begin{verbatim}
item = generate_item_bank(1)[0]
plot.item_curve(item[0], item[1], item[2], item[3], ptype='icc')
plot.item_curve(item[0], item[1], item[2], item[3], ptype='iic')
plot.item_curve(item[0], item[1], item[2], item[3], ptype='both')
\end{verbatim}

\begin{figure}
	\centering
	\includegraphics[width=.7\textwidth]{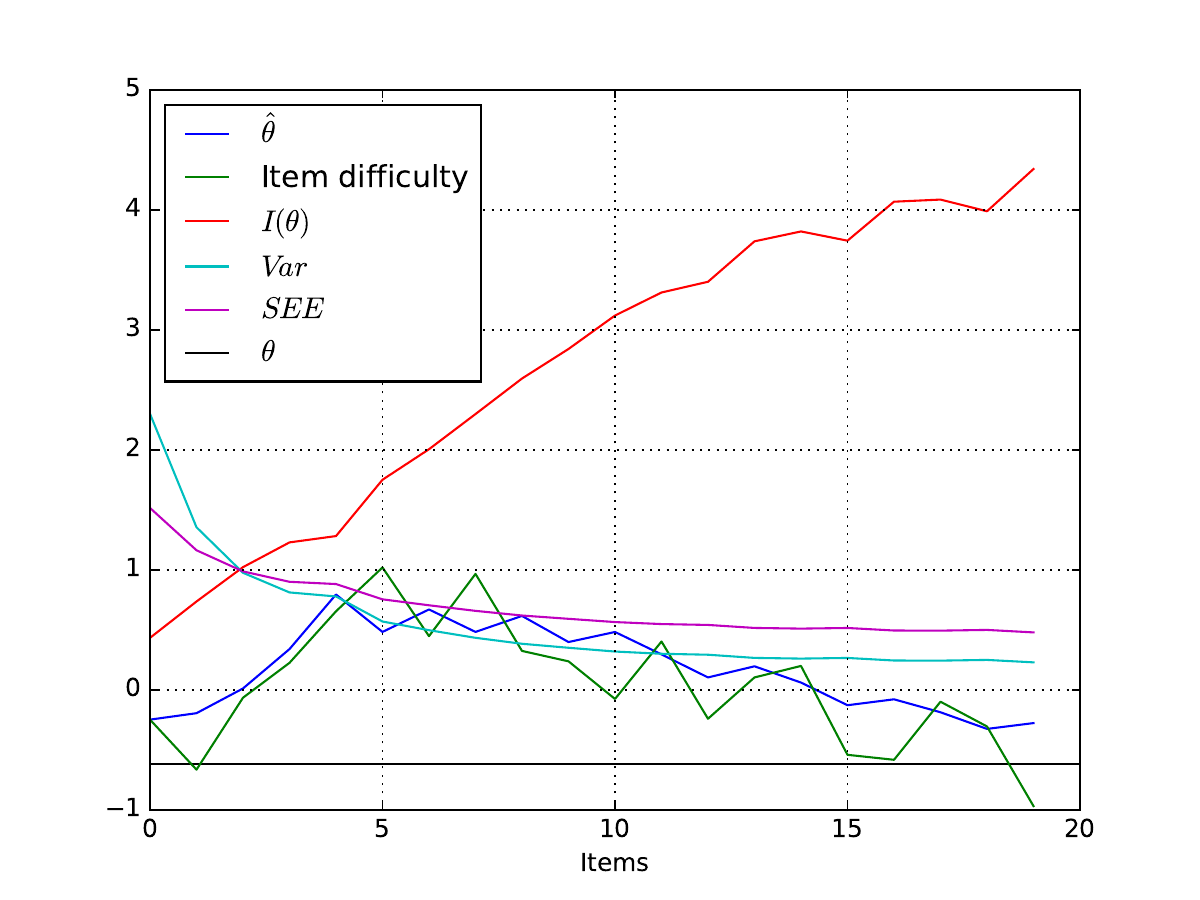}
	\caption{Graphics showing the progress of certain measurements during an examinee's test.}\label{fig:progress}
\end{figure}

Given a simulator \code{s}, the test progress of an examinee can be plotted using the function \code{test_progress()} and passing the examinee's index, like so:

\begin{verbatim}
plot.test_progress(simulator=s, index=0, true_theta=s.examinees[0],
    info=True, var=True, see=True)
\end{verbatim}

These commands generate the graphics depicted in figure \ref{fig:progress}.

\begin{figure}[!ht]
	\centering
	\subfloat[Histogram]{\includegraphics[width= 0.49\columnwidth]{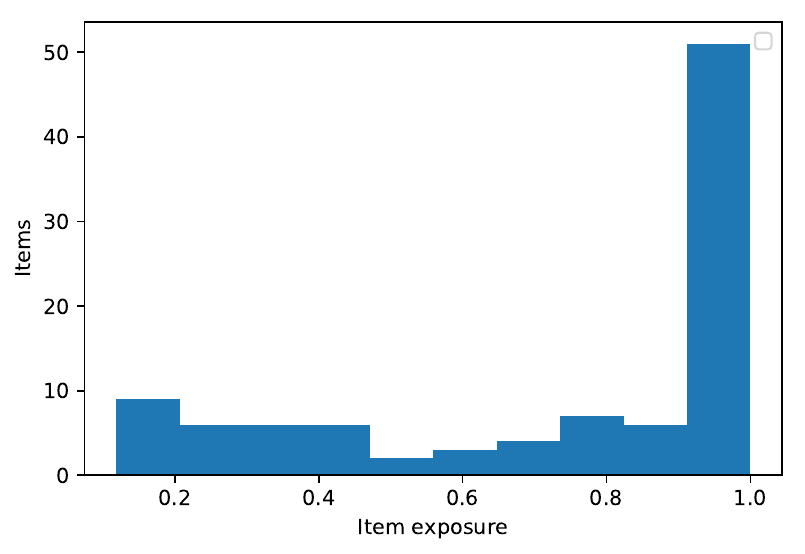}}
	\subfloat[Dotted line chart]{\includegraphics[width= 0.49\columnwidth]{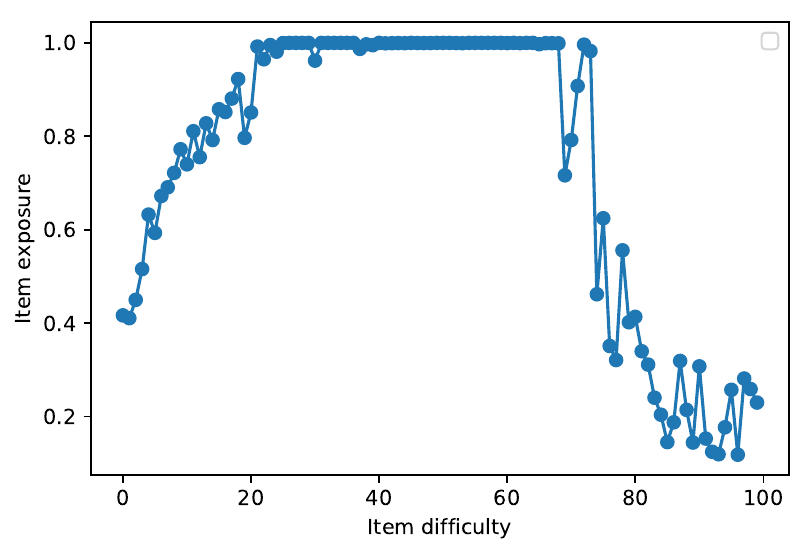}}
	\caption{Charts presenting item exposure rates of an item bank.}\label{fig:exposures}
\end{figure}

At last, item exposure rates can be plotted via the \code{item_exposure()} function. Given a simulator \code{s} or an item matrix \code{i} with the last column representing the items exposure rate, item exposure can be plotted as the following example shows:

\begin{verbatim}
plot.item_exposure(simulator=s, hist=True)
plot.item_exposure(simulator=s, hist=False, par='b')
\end{verbatim}

Both of these commands generate graphics similar to the one shown in figure \ref{fig:exposures}. The \code{par} parameter allows exposure rates to be sorted by one of the item parameters before plotting. The \code{ptype} parameter allow the chart to be plotted using either bars or lines.

\section{Examples}

\begin{figure}
	\centering
	\includegraphics[width=0.7\textwidth]{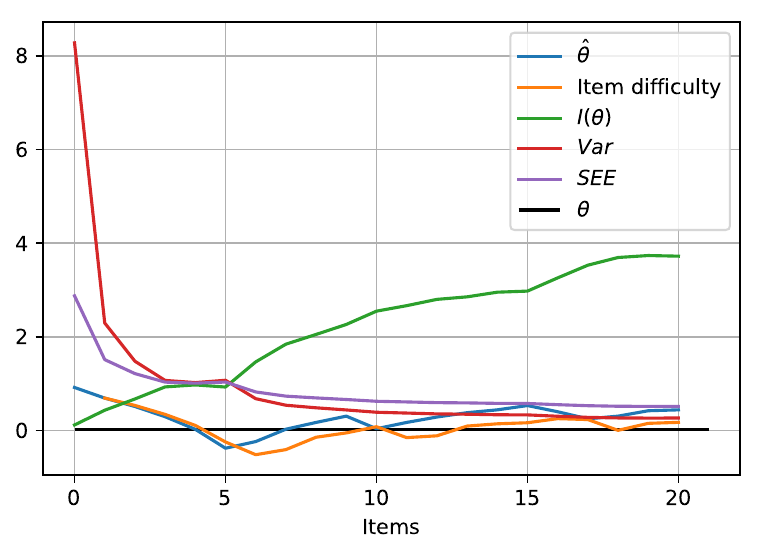}
	\caption{CAT results of an examinee. Items were selected by the maximum information selector and test stops at 20 items.}
	\label{fig:example1}
\end{figure}

This section presents different examples of running simulations through \pkg{catsim}. Figure \ref{fig:example1} presents a simulation with 10 examinees, a random \(\hat\theta_0\) initializer, a maximum information item selector, a hill climbing estimator and a stopping criterion of 20 items.

\begin{verbatim}
s = Simulator(items, 10)
s.simulate(RandomInitializer(), MaxInfoSelector(),
    HillClimbingEstimator(), MaxItemStopper(20))
catplot.test_progress(simulator=s,index=0)
\end{verbatim}

\begin{figure}
	\centering
	\includegraphics[width=0.7\textwidth]{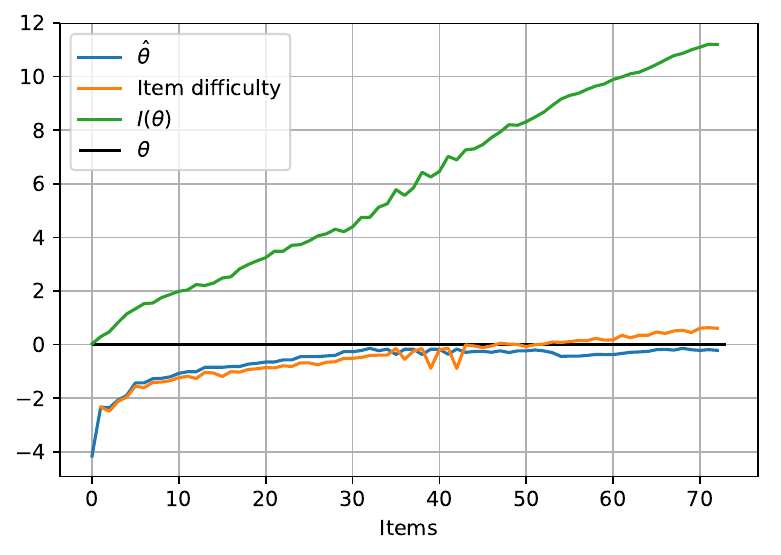}
	\caption{CAT results of an examinee. Items were selected by the maximum information selector and test stops when a minimum error threshold is reached.}
	\label{fig:example2}
\end{figure}

Figure \ref{fig:example2} presents a way of using predetermined proficiencies by passing either a \code{list} or a one-dimensional \code{numpy.ndarray}. It also uses the minimum error stopping criterion, making each examinee have tests of different sizes.

\begin{verbatim}
examinees = numpy.random.normal(size=10)
s = Simulator(items, examinees)
s.simulate(RandomInitializer(), MaxInfoSelector(),
    HillClimbingEstimator(), MinErrorStopper(.3))
catplot.test_progress(simulator=s,index=0, info=True)
\end{verbatim}

The last example shows a linear item selector, whose item indexes are selected in order from a \code{list}. The example shows indexes being randomly selected from the item bank, but they can also be passed directly, such as \code{indexes = [1, 10, 12, 27, 116]}.

\begin{verbatim}
s = Simulator(items, 10)
indexes = numpy.random.choice(items.shape[0], 50, replace=False)
s.simulate(RandomInitializer(), LinearSelector(indexes),
    HillClimbingEstimator(), MaxItemStopper(50))
catplot.test_progress(simulator=s,index=0, info=True, see=True)
\end{verbatim}

\begin{figure}
	\centering
	\includegraphics[width=0.7\textwidth]{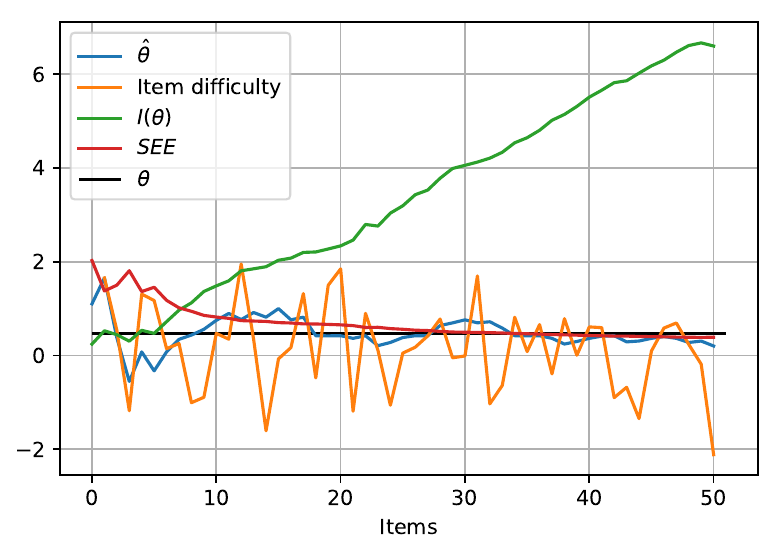}
	\caption{Test results of an examinee. Items were selected linearly and test stops at 50 items.}
	\label{fig:example3}
\end{figure}

\section{Discussion}

This paper has presented \pkg{catsim}, a computerized adaptive testing package written in \proglang{Python}. \pkg{catsim} separates each phase of the CAT iterative process into different components and provides implementations of different methods from the literature, as well as interfaces which can be used by end users to create their own methods. A simulator provides a complete history of the test progress for each examinee, including items chosen, answers given and measures taken after each answer, as well as validity measures for the whole test. Simulation results, item parameters and exposure rates can then be plotted with the provided functions. Individual components of \pkg{catsim} can also be used independent of simulations, enabling users to incorporate the package in third-party test-taking applications.

\pkg{catsim} fills a gap in the \proglang{Python} scientific community, being the only package specialized in computerized adaptive tests and the logistic models of Item Response Theory. Other functionalities that may benefit the community include item parameter and examinee proficiency estimation through a dichotomous data matrix and the implementation of Bayesian estimators, whose behavior contrasts with the maximum likelihood estimators already provided.

\bibliography{referencias}

\end{document}